\documentclass[11pt,english]{article}
\usepackage[utf8]{inputenc}
\usepackage{amsmath}
\usepackage{amsthm}
\usepackage{amssymb}
\usepackage{graphicx}

\makeatletter
\@ifundefined{date}{}{\date{}}
\usepackage{t1enc}
\usepackage[english]{babel}
\usepackage{amsthm}
\usepackage{braket}
\usepackage{centernot}
\usepackage{tikz}

\def\cros{\raise1.9pt\hbox{$\scriptscriptstyle
          >$}\!\raise1.5pt\hbox{$\scriptstyle\triangleleft\,$}}

\theoremstyle{definition}\theoremstyle{definition}\theoremstyle{definition}\theoremstyle{definition}\newcommand{\noi}{\vspace{0.1in} \noindent}

\title{\bf PBR, nonreality and entangled measurement}
\author{\textit{Gábor Hofer-Szabó}\thanks{HUN-REN Research Center for the Humanities, Budapest, email: szabo.gabor@abtk.hu}}

\makeatother

\usepackage{babel}
\begin{document}
\maketitle
\begin{abstract}
In a recent paper, Cabbolet argues that the PBR theorem is nonreal since in the ensemble interpretation of quantum mechanics the entangled measurement used in the derivation of the PBR theorem is nonexisting. However, Cabbolet (1) doesn't provide any argument for the nonexistence of entangled measurements beyond the incompatibility of the existence of entangled measurements and the existence of $\psi$-epistemic models which we already know from the PBR theorem; and (2) he doesn't show why it is more reasonable to abandon  entangled measurements instead of $\psi$-epistemic models. Hence, the PBR theorem remains intact. 
\vspace{0.1in}

\end{abstract}

The question of what exactly Einstein's  ensemble  interpretation of quantum mechanics (QM) consisted of is a delicate question \cite{Fine}. Nevertheless, in the modern framework of operational theories and ontological  models \cite{Spekkens}, the Einsteinian view has become to be identified as a $\psi$-epistemic ontological model for QM \cite{Harrigan}. The $\psi$-epistemic models, however, are ruled out by the PBR theorem \cite{PBR}. But  PBR's argument has an implicit  assumption: the existence of entangled  measurements.  Thus, the PBR theorem can be read as an argument establishing a \textit{contradiction} between the  existence of $\psi$-epistemic models and the  existence of entangled  measurements. PBR resolves the contradiction by abandoning $\psi$-epistemic models. This  is a plausible choice since we have no direct empirically access to ontological models while routinely apply entangled measurements  in QM.

In a recent paper \cite{Cabbolet}, Cabbolet chooses the opposite way out of the contradiction. He explicitly constructs a classical model mimicking PBR's bipartite ensembles. The model is "$\psi$-epistemic" in the sense that there are individual systems in each ensemble that share two properties mimicking PBR's two preparations, $\ket{0}$ and $\ket{+}$. Cabbolet then argues that there is no entangled measurement for this model. Consequently, the PBR theorem is nonreal since in $\psi$-epistemic  models entangled measurements are nonexisting. 

Cabbolet's argument consist in the construction of a specific "entangled measurement" on his classical ensembles and showing that this measurement will not yield an outcome for a $1/4$ of the cases. Consequently,  this measurement cannot exist. This is a strange way of proving nonexistence. One  usually doesn't  prove $\lnot \exists  x.Px$  by showing that $\lnot Pa$ for a given $a$. But the important point is this: Cabbolet's example is \textit{unnecessary} since he need not prove in any way that the existence of $\psi$-epistemic models rule out the existence of entangled  measurements. We know that from the PBR theorem. The PBR theorem is just a contradiction between the existence of $\psi$-epistemic models and  entangled  measurements.  \textit{What he needs to show is why it is more rational to abandon entangled  measurements instead of $\psi$-epistemic models}. To this question, however, Cabbolet does not give an answer. 

Cabbolet's strategy against PBR is a kind of "your \textit{modus ponens} is my \textit{modus tollens}".  PBR takes it for granted that entangled  measurements do exist and use the theorem to argue against $\psi$-epistemic models for QM. In contrast, Cabbolet explicitly constructs a classical $\psi$-epistemic toy model and argues that there can exist no entangled  measurements. But the question of the existence of entangled  measurements is a factual 	question. Experimenters routinely perform measurements in different entangled bases. On the other hand, Cabbolet's  toy model is not a full-fledged $\psi$-epistemic model for QM. So the decision between abandoning epistemic models or entangled measurements is not as symmetric as it seems from the pure logic of the PBR argument. 

The main problems with Cabbolet's paper is that (1) it doesn't provide any argument for the nonexistence of entangled measurements beyond the incompatibility of the existence of entangled measurements and $\psi$-epistemic models which we already know from the PBR theorem; and (2) it doesn't show why it is more reasonable to abandon  entangled measurements instead of $\psi$-epistemic models. Therefore, I think the PBR theorem and its usual interpretation remains intact. 

I also have some other problems with Cabbolet's paper. Namely, his "entangled measurement" (i) is not the entangled measurement of PBR and (ii) from its nonexistence it does not follow that there are no entangled measurements in general. To tackle  these problems, I need to go deeper into Cabbolet's paper. But (i)-(ii) are only side problems since they all concern Cabbolet's specific strategy to prove that entangled measurements are nonexisting. But, as I said above, to show this Cabbolet should provide a $\psi$-epistemic model for QM. So my main criticism is problem (1) and (2) above.

\noi
Let us start with some terminology. An \textit{ontological (hidden variable) model} identifies the quantum state with a preparation procedure generating an ensemble of systems each in a given ontic (hidden) states $\lambda$.  The model fixes the ontology as a set of ontic states $\Lambda =\{\lambda\}$ and associates with every (pure) quantum state $\ket{\psi}$ a probability distribution $\mu_\psi(\lambda)$ of ontic states. If these ontic states uniquely determine the outcome of every measurement, then the ontological model is called \textit{outcome-deterministic} \cite{Spekkens}. If the support of the probability distribution $\mu_\psi(\lambda)$ and  $\mu_{\psi'}(\lambda)$ of two different quantum states $\ket{\psi}$ and  $\ket{\psi'}$ can overlap, then we call the model \textit{$\psi$-epistemic}, otherwise \textit{$\psi$-ontic} \cite{Harrigan}. The PBR theorem shows that QM does not allow for a $\psi$-epistemic ontological model. 

The argument runs as follows. In Step 1, consider two quantum states  (preparation procedures) $\ket{0}$ and $\ket{+} = \frac{1}{\sqrt{2}}(\ket{0} + \ket{1})$ and suppose \textit{ad absurdum} that the support of the associated probability distributions $\mu_0(\lambda)$ and $\mu_+(\lambda)$ overlap. Next, consider four ensembles each consisting of independently prepared \textit{pairs} of systems corresponding to the following four quantum states:
\begin{eqnarray*}
\ket{0} \otimes \ket{0} \qquad \ket{0} \otimes \ket{+} \qquad \ket{+} \otimes \ket{0} \qquad \ket{+} \otimes \ket{+} 
\end{eqnarray*}
Since the probability distributions $\mu_0(\lambda)$ and $\mu_+(\lambda)$ overlap, there will be a pair in each of the four ensembles with the same ontic state $\lambda$ lying in the overlap region. 

Now, suppose we perform an entangled measurement on  this very pair of systems. The outcomes of this measurement are represented in QM by the following four entangled states: 
\begin{eqnarray*}
\ket{\xi_1} &=& \frac{1}{\sqrt{2}}\big(\ket{0} \otimes \ket{1} +\ket{1} \otimes \ket{0} \big) \\
\ket{\xi_2} &=& \frac{1}{\sqrt{2}}\big(\ket{0} \otimes \ket{-} +\ket{1} \otimes \ket{+} \big) \\
\ket{\xi_3} &=& \frac{1}{\sqrt{2}}\big(\ket{+} \otimes \ket{1} +\ket{-} \otimes \ket{0} \big) \\
\ket{\xi_4} &=& \frac{1}{\sqrt{2}}\big(\ket{+} \otimes \ket{-} +\ket{-} \otimes \ket{+} \big) 
\end{eqnarray*}
where $\{\ket{0}, \ket{1}\}$ is a basis in the qubit space and $\ket{-} = \frac{1}{\sqrt{2}}(\ket{0} - \ket{1})$. Since the entangled states "antidistinguish" the preparations (Leifer, 2014), that is each entangled state $\ket{\xi_i}$ is orthogonal to one of the above quantum states
\begin{eqnarray*}
\ket{\xi_1}  &\perp & \ket{0} \otimes \ket{0} \\
\ket{\xi_2}  &\perp & \ket{0} \otimes \ket{+} \\
\ket{\xi_3}  &\perp & \ket{+} \otimes \ket{0} \\
\ket{\xi_4}  &\perp & \ket{+} \otimes \ket{+} 
\end{eqnarray*}
therefore our pair of systems cannot yield result 1 when prepared in the state $\ket{0} \otimes \ket{0}$, cannot yield result 2 when prepared in the state $\ket{0} \otimes \ket{+}$, and so on. But the outcome of the measurement cannot depend on which procedure the pair was previously prepared by; it can depend only on the ontic state of the pair. (This assumption is called "$\lambda$-sufficiency" by Spekkens.) Thus, the measurement cannot yield an outcome. Contradiction. Hence, $\mu_0(\lambda)$ and $\mu_+(\lambda)$ cannot overlap. In Step 2, PBR derives a similar contradiction using multiple tensor products for \textit{any} two distinct quantum states, hence ruling out a $\psi$-epistemic interpretation of QM in general.

Now, I turn to Cabbolet's paper.

Cabbolet mimics Step 1 of PBR's bipartite ensembles  by  two large batches of bolts. Each bolt can be of type  M12 or type M10 and can be either sellable or unsellable. We perform measurements on the bolts. Let  $A$ denote the measurement of the type of the bolts with outcomes $0$ (type M12) and $1$ (type M10) and let  $B$ (in Cabbolet: measurement $P$) denote the  measurement of the sellability of the bolts with outcomes $+$ (sellable) and $-$ (unsellable). I assume that Cabbolet's example with the bolts should not be read \textit{literally}. In this case, namely, one would not be surprised that there is no entangled measurement which can be performed on such classical systems as bolts. This reading is reinforced also by the fact that in the description of the measurement statistics Cabbolet applies the formalism of QM. Thus, I will take Cabbolet's  model with the bolts simply \textit{metaphorically} as an illustration of a part of a $\psi$-epistemic ontological model of QM and I take the measurements $A$ and $B$ simply as measurements on a quantum system, disregarding the specific details of his example.

Consider then two ensembles of quantum systems, one prepared in the state  $\ket{0}$, the other prepared in the state $\ket{+}$. Suppose we perform two measurement $A$ and $B$ on the systems which are represented in QM by the Pauli matrices $\sigma_z$  and $\sigma_x$. If we perform measurement $B$ on the first ensemble, we obtain the outcome $+$ in half of the cases and $-$ in the other half. Similarly, if we perform a measurement $A$ on the second ensemble, we obtain the outcome $0$ in half of the cases and $1$ in the other half. Denote by $\Lambda_{0+}$ the set of ontic states for which measurement $A$ would definitely yield outcome $0$ and measurement $B$ would definitely yield outcome $+$.\footnote{Cabbolet introduces the quantum state $\ket{0+}$ to represent the ensemble $\Lambda_{0+}$. But such an ensemble cannot be prepared quantum mechanically; so it is better to refer to it at the level of the ontological model as $\Lambda_{0+}$.} Similarly define $\Lambda_{0-}$, $\Lambda_{1+}$ and $\Lambda_{1-}$ for the other outcome pairs. (Note that $A$ and $B$ need not be performed simultaneously for $\Lambda_{0+}$ to exist.) Then, in our first ensemble the half of the ontic states is in $\Lambda_{0+}$ and the other half is in $\Lambda_{0-}$. In our second ensemble the half of the ontic states is in $\Lambda_{0+}$ and the other half is in $\Lambda_{1+}$.

Now, consider again the four ensembles each consisting of independently prepared pairs of systems corresponding to the above four quantum states:
\begin{eqnarray*}
\ket{0} \otimes \ket{0} \qquad \ket{0} \otimes \ket{+} \qquad \ket{+} \otimes \ket{0} \qquad \ket{+} \otimes \ket{+} 
\end{eqnarray*}
It follows that in each of the four ensembles the proportion of those pairs which are in an ontic state $(\lambda,\lambda')\in \Lambda_{0+} \times \Lambda_{0+}$ will be $1/4$. 

Cabbolet claims that for these pairs the entangled measurement represented by the entangled states $\{\ket{\xi_i}\}$ will not give an outcome. This is true but not new; it follows simply from the PBR theorem since $\Lambda_{0+}$ is in the intersection of the support of $\mu_0(\lambda)$ and $\mu_+(\lambda)$. At this point, however, Cabbolet and PBR, as said before, come to different conclusions. PBR argues that because the  entangled measurement would not give an outcome, therefore there is no  $\psi$-epistemic ontological model for QM. In contrast, Cabbolet argues that because the entangled measurement would not give an outcome, therefore there are no such entangled measurements. Both parties agree that $\psi$-epistemic ontological  models and entangled measurement represented by the states $\{\ket{\xi_i}\}$ are incompatible but they differ in which assumption to give up.

To decide the dispute in favour of Cabbolet, one would like to see a full-fledged $\psi$-epistemic model of QM with the ontic regions $\Lambda_{0+}$, $\Lambda_{0-}$, $\Lambda_{1+}$ and $\Lambda_{1-}$ embedded into the model. However, we don't have such a model (Leifer, 2014, 72) and neither Cabbolet offers one. On the other hand, experimental techniques are rapidly evolving today and entangled measurements are applied in experimental physics on a daily basis. We stress here that the \textit{sole criterion} of whether a measurement qualifies as an entangled measurement represented by the states $\{\ket{\xi_i}\}$ is that it should yield the outcomes  in \textit{every} quantum state $\ket{\psi}$ with probability  $|\!\braket{\psi,\xi_i}\!|^2$. The existence of entangled measurement doesn't in any sense depend on whether certain ontological models of QM are possible or not. All this tips the dispute in favor of PBR's interpretation of the PBR theorem: one should give up $\psi$-epistemicity and retain entangled measurements. 

Now, I turn to the smaller problems (i) and (ii).

First, let's see  how Cabbolet realizes the  entangled measurement $\{\ket{\xi_i}\}$.  He considers the following four measurement instructions: 
\begin{description}
\item M$_1$: Perform measurement $A$ on both ensembles;
\item M$_2$: Perform measurement $A$ on the first ensembles and measurement $B$ on the second;
\item M$_3$: Perform measurement $B$ on the first ensembles and measurement $A$ on the second;
\item M$_4$: Perform measurement $B$ on both ensembles.
\end{description}
Cabbolet then associates each \textit{measurement} M$_i$ with the entangled state $\ket{\xi_i}$. But this is no good. Each measurement M$_i$ has four outcomes. Which of the four  outcomes of the four measurements should be associated with which entangled state? Here is one way to make Cabbolet's idea precise:
\begin{description}
\item M$_1'$: Perform measurement $A$ on both ensembles and write $+1$ if the outcomes are opposite and write $-1$ if they are the same;
\item M$_2'$: Perform measurement $A$ on the first ensembles and measurement $B$ on the second and 
write $+1$ if the outcomes are $(0,-)$ or $(1,+)$ and write $-1$ if they are $(0,+)$ or $(1,-)$;
\item M$_3'$: Perform measurement $B$ on the first ensembles and measurement $A$ on the second and 
write $+1$ if the outcomes are $(+,1)$ or $(-,0)$ and write $-1$ if they are $(+,0)$ or $(-,1)$;
\item M$_4'$: Perform measurement $B$ on both ensembles and write $+1$ if the outcomes are opposite and write $-1$ if they are the same.
\end{description}
So in fact Cabbolet associates not the measurement M$_i'$ but the \textit{outcome} $+1$ of M$_i'$ with the entangled state $\ket{\xi_i}$.  Then he  shows that for the pairs in ontic state $(\lambda,\lambda')\in \Lambda_{0+} \times \Lambda_{0+}$ none of the four M$_i'$ will provide outcome $+1$. Thus, the entangled measurement $\{\ket{\xi_i}\}$  has no outcome. 

But why does the outcome $+1$ of M$_i'$ correspond to the entangled state $\ket{\xi_i}$? Cabolet's answer is that because if we prepare the two batches of bolts not independently but putting the pairs of bolts in pairs of envelopes in a certain way specified in the paper, then all the measurements M$_i'$ will give outcome $+1$. I am unable to see how this fact justifies that the outcome $+1$ of the measurement M$_i'$ corresponds to the entangled state $\ket{\xi_i}$. In the formalism of QM, the outcome $+1$ of the measurement M$_1'$ corresponds to coarse-graining over those pairs of outcomes of a measurement with basis vectors 
\begin{eqnarray*}
\big\{\ket{0} \otimes \ket{0}, \ket{0} \otimes \ket{1} , \ket{1} \otimes \ket{0}, \ket{1} \otimes \ket{1} \big\}
\end{eqnarray*}
for which the product is $+1$; while $\ket{\xi_i}$ is the entangled basis vector
\begin{eqnarray*}
\frac{1}{\sqrt{2}}\big(\ket{0} \otimes \ket{1} +\ket{1} \otimes \ket{0} \big)
\end{eqnarray*}
They are definitely not the same. 

More generally, entangled measurements, like the Bell state measurement for example, are typically complicated \textit{global} measurements  which cannot be reduced to \textit{local} measurements of the two parties. Measuring a photon pair in a Bell basis requires recombining the two beams by a beam splitter and not just performing polarization measurements on the two distant photons \cite{Weihs}. Anyway, we can easily check whether  M$_1'$-M$_4'$ is an entangled measurement. If it is, then the outcome statistics in every quantum state needs to be calculated by the Born rule, that is taking the square of the product of the four entangled state $\ket{\xi_i}$ with the quantum state. Pick, for example,  the quantum state (preparation) $\ket{0} \otimes \ket{1}$. What is the probability to  obtain the outcome $+1$ for the measurements M$_1'$  in this state? The probability is $1$ since measuring $A$ on both sides, we always get opposite outcomes. But this is not the same as $|\!\bra{\xi_1}(\ket{0} \otimes \ket{1})\!|^2$ which is $1/2$. So the $+1$ outcomes of M$_1'$-M$_4'$ cannot correspond to the states $\{\ket{\xi_i}\}$ of an entangled measurement.

But regardless of M$_1'$-M$_4'$, can we not show directly that if we perform a "true" entangled measurement $\{\ket{\xi_i}\}$ on a pair in state $(\lambda,\lambda')\in \Lambda_{0+} \times \Lambda_{0+}$, then we will get no outcome? Yes, we can. This fact follows directly from the PBR theorem. But again, to show that the entangled measurement $\{\ket{\xi_i}\}$ does not yield an outcome and hence it is nonexisting, we should first know that QM has a  $\psi$-epistemic model with some $(\lambda,\lambda')\in \Lambda_{0+} \times \Lambda_{0+}$. But we don't know that.

In sum, Cabbolet devises an entangled measurement and shows that this measurement does not yield an outcome for a $1/4$ of the cases, therefore it is nonexistent. Next, he concludes that there exist no entangled measurements in general. But Cabbolet's "entangled measurement" (i) is  not the entangled measurement $\{\ket{\xi_i}\}$ and (ii) from its nonexistence it does not follow that there are no entangled measurements in general. To show that there are no entangled measurements one should first give a full-fledged $\psi$-epistemic ontological models for QM and use the PBR theorem in a reversed way. Until we have such model, it is more reasonable to draw the standard
conclusion from the incompatibility of the PBR, namely that it  rules out $\psi$-epistemic ontological models for QM.

\end{document}